\documentstyle[preprint,prl,aps]{revtex}

\title{ Chemical freezeout in relativistic $A$+$A$ collisions: \\
is it close to the QGP? }

\author{ Mark I. Gorenstein,$^{a,b,d,}$\footnote{Permanent address: 
Bogolyubov Institute for Theoretical Physics, Kiev, Ukraine } 
Horst St\"ocker,$^{a,b}$ Granddon D. Yen,$^c$ Shin Nan Yang,$^d$ \\ 
and Walter Greiner$^a$ }

\address{ $^a$Institute for Theoretical Physics, Goethe University, 
Frankfurt, Germany \\
$^b$Department of Physics and Astronomy, TAU, Tel Aviv, Israel\\
$^c$Institute of Physics, Academia Sinica, Taipei 11529, Taiwan \\
$^d$Department of Physics, National Taiwan University, Taipei 10617, 
Taiwan }

\draft

\begin{document}

\maketitle

\begin{abstract}
Preliminary experimental data for particle number ratios in the 
collisions of Au+Au at the BNL AGS (11$A$ GeV/$c$) and Pb+Pb at the 
CERN SPS (160$A$ GeV/$c$) are analyzed in a thermodynamically 
consistent hadron gas model with excluded volume.  Large values of 
temperature, $T=140$--185 MeV, and baryonic chemical potential, 
$\mu_b=590$--270 MeV, close to the boundary of the quark-gluon plasma 
phase are found from fitting the data.  This seems to indicate that 
the energy density at the chemical freezeout is tremendous which 
would be indeed the case for the point-like hadrons.  However, a 
self-consistent treatment of the van der Waals excluded volume 
reveals much smaller energy densities which are very far below a 
lowest limit estimate of the quark-gluon plasma energy density.  
\end{abstract}



The quantitative results of quantum chromodynamics for the equation 
of state of strongly interacting matter have been obtained mostly in 
the high energy density region, $\varepsilon>2$~GeV/fm$^3$, where 
one expects a weakly interacting gas of quarks and gluons --- the 
quark-gluon plasma (QGP).  The region of small $\varepsilon$ 
corresponds to the hadron gas (HG) phase and can only be treated 
today in terms of the hadron degrees of freedom.  Assuming a local 
thermodynamical equilibrium at the final freeze-out stage of 
nucleus-nucleus ($A$+$A$) collisions, one can estimate the particle 
number ratios without detailed knowledge of the complicated system 
evolution.  Hadron abundances and ratios have been suggested as 
possible signatures for exotic states and phase transitions in dense 
nuclear matter (see, e.g., Refs.~[1--5]).  

Preliminary data for $A$+$A$ collisions with truly heavy beams have 
recently become available: Au+Au at 11$A$ GeV/$c$ at the BNL AGS and 
Pb+Pb at 160$A$ GeV/$c$ at the CERN SPS \cite{qm96}.  A systematic 
analysis of these data could yield clues to whether a short-lived 
QGP phase with high $\varepsilon$ exists during the hot and dense 
stage of these reactions.  Recent fits of particle number 
ratios [7--12] at AGS and SPS energies in the framework of ``thermal 
model'' have led to large values of temperature, $T$, and baryonic 
chemical potential, $\mu_b$, close to the boundary of the QGP 
phase \cite{braun95b,stachel}.  The aim of the present paper is 
two-fold.  First, we clarify the notion of the ``thermal model'' 
used in the literature.  Second, we present our results for Au+Au 
(AGS) and Pb+Pb (SPS) collisions and examine whether the chemical 
freeze-out states are really close to the QGP.  We find large values 
of the chemical freeze-out parameters $T=140$--185 MeV and 
$\mu_b=590$--270 MeV.  The values of the energy density $\varepsilon$ 
and baryonic number density $n_b$ are, however, strongly dependent on 
the specific thermal model formulation.  In the ideal HG, the 
chemical freeze-out values of $\varepsilon$ and $n_b$ for SPS $A$+$A$ 
collisions are indeed close to the QGP estimates.  These values 
become, however, much smaller in thermodynamically consistent HG 
model with excluded volume.  

The name ``thermal model'' has always been used in the literature 
when calculations of particle number ratios in $A$+$A$ collisions are 
done with parameters $T$ and $\mu_b$.  We stress, however, that 
different people used, in fact, very different ``thermal models''.  
It seems natural to use the ideal HG as a thermal model at the 
freeze-out stage.  However, such an ideal gas model becomes 
inadequate in high-energy $A$+$A$ collisions.  The chemical 
freeze-out parameters $T$ and $\mu_b$ obtained from fitting the 
particle number ratios at AGS and especially SPS energies lead to 
artificially high particle number densities which contradict the 
assumed picture of non-interacting hadrons.  The van der Waals (VDW) 
excluded volume procedure has recently been used in a number of 
thermal model calculations.  We follow the excluded volume procedure 
of Ref.~\cite{gor91}.  The ``thermal models'' of Refs.~[7--12] also 
include VDW ``corrections'', but in {\it ad hoc} thermodynamically 
inconsistent ways.  Besides, in the formulation of Refs.~[7--10],  
particle ratios always remain the same as in the ideal gas for any 
choice of proper hadron volumes.  

For a fixed particle number $N$, the VDW excluded volume procedure 
amounts to the substitution of the volume $V$ by $V-vN$, where $v$ is 
the parameter corresponding to the proper volume of the particle.  
Note that this VDW procedure, interpreted in statistical mechanics as 
an approximation to  the gas of hard-sphere particles with radius 
$r$, requires that the volume parameter $v$ is equal to the 
``hard-core particle volume'', $\frac{4}{3}\pi r^3$, multiplied by a 
factor of 4 \cite{lan75}.  An extension of the excluded volume 
procedure for several particle species $i=1,...,h$, has been done 
with the {bf ansatz of the} substitution 
$V~\rightarrow ~V-\sum_{i=1}^{h} v_i N_i~$.  The pressure function in 
the grand canonical formulation is given then by the 
equation \cite{gor91}: 
\begin{equation}\label{vdweos}
p(T,\mu_1,..., \mu_h)~ = ~
\sum_{i=1}^h p_i^{id}(T,~\mu_i~ - ~ v_i ~ p(T,\mu_1,...,\mu_h))~,
\end{equation}
where $p_i^{id}$ is the ideal gas pressure of $i$-th particle 
$$p_i^{id}(T,\mu_i)~= ~\frac{d_i}{6\pi^2}\int_0^{\infty}k^2dk~
\frac{k^2}{(k^2+m_i^2)^{1/2}}
\left[\exp\left(\frac{(k^2+m_i^2)^{1/2}-\mu_i}
{T}\right) \pm 1\right]^{-1}~,$$ 
with $d_i$ the number of internal degrees of freedom (degeneracy), 
$\mu_i$ the particle chemical potential and $m_i$ the mass.  Particle 
number density for the $i$-th species is calculated from 
Eq.~(\ref{vdweos}) as 
\begin{equation}\label{partden}
n_i~\equiv~ \frac{\partial p}{\partial \mu_i}
=~\frac{n_i^{id}(T, \tilde{\mu}_i)}
{1~+~\sum_{j=1}^h
v_j~n_j^{id}(T, \tilde{\mu}_i)}~,
\end{equation}
with 
\begin{equation}\label{muitilde}
\tilde{\mu}_i~ \equiv~ \mu_i ~ - ~v_i~ p(T,\mu_1,..., \mu_h)~,~~~
i=1,...,h~.  
\end{equation}
The energy density of the HG with VDW repulsion is 
\begin{equation}\label{epsdenh}
\varepsilon ~
=~\frac{\sum_{i=1}^{h}\varepsilon_i^{id}(T,
\tilde{\mu}_i)}
{1~+~\sum_{j=1}^h
v_j~n_j^{id}(T, \tilde{\mu}_j)}~,
\end{equation}
where $n_i^{id},~\varepsilon_i^{id}$ are the ideal gas quantities.  
The VDW repulsion leads to strong suppression of particle number 
densities and therefore all thermodynamical functions in the HG with 
the VDW repulsion become much {\it smaller} than those in the ideal 
gas at the same $T$ and $\mu_1,...,\mu_h$ (see Ref.~\cite{yen97} for 
details).  

Particle chemical potentials $\mu_i$ regulate the values of conserved 
charges.  For simplicity we neglect the effects of non-zero 
electrical chemical potential which were considered in 
Ref.~\cite{goryang91,cleym97}.  The chemical potential of the $i$-th 
particle can then be written as 
\begin{equation}\label{mu}
\mu_i ~=~ b_i\,\mu_b~+~s_i\,\mu_s~, 
\end{equation}
in terms of baryonic chemical potential $\mu_b$ and strange chemical 
potential $\mu_s$, where $b_i$ and $s_i$ are the corresponding 
baryonic number and strangeness of the $i$-th particle.  The hadronic 
gas state is defined by two independent thermodynamical parameters, 
$T$ and $\mu_b$.  The strange chemical potential $\mu_s(T,\mu_{b})$ 
is determined from the requirement of zero strangeness 
\begin{equation}\label{strange}
n_{s}(T,\mu_b,\mu_{s})~\equiv~\sum_{i=1}^{h}
s_{i}n_{i}(T,\mu_{i})~=~0~.
\end{equation}

For the SPS $A$+$A$ data it has been observed in 
Refs.~\cite{cleym93b,cleym94} that their ``thermal model'' (for 
particle number ratios it is, in fact, equivalent to the ideal gas 
model) is unable, within a single set of freeze-out parameters, to 
reproduce simultaneously the strange particles and anti-baryon to 
baryon ratios together with ratios where pions are involved.  
Experimental pion to nucleon ratio and ratios of pions to other 
hadrons are larger than the ideal gas predictions.  Several 
mechanisms, including the chemical non-equilibrium with $\mu_{\pi}>0$ 
in the pion subsystem and the possibility of straight QGP 
hadronization \cite{cleym93b,cleym94,let93} were proposed, but no 
satisfactory solution has been found.  Another possibility with an 
incomplete strangeness chemical equilibrium and parameter 
$\gamma_s<1$ has been recently studied in Ref.~\cite{bec97}.  We 
follow the HG model with complete chemical equilibrium.  To enlarge 
the number of pions in our VDW HG we assume that the pion proper 
volume parameter $v_{\pi}$ is smaller than the proper volume 
parameters for other hadrons which we put to be equal to $v$.  
In the case where not all of the $v_i$'s are equal, the hadron volume 
parameters $v_{i}$ do influence	the particle number ratios through 
the modification of Eq.~(\ref{muitilde}) in the particle chemical 
potentials, required by thermodynamical self-consistency.  The effect 
is quite evident: hadrons which take up less space (i.e., smaller 
values of $v_{i}$) are preferably produced.  The particle number 
ratios of those small hadrons to larger ones increase as compared 
with the ideal gas results.  We assume different hard-core radii: 
$r_{\pi}$ for pions and $r$ for all other hadrons ($r>r_{\pi}$), with 
$v_i = 4\cdot {4\pi\over 3}~ r_i^3$.  

Baryon and meson resonances are of great importance for the measured 
particle number ratios at AGS and SPS energies.  All known resonance 
states with masses up to 2 GeV are included in our calculations with 
subsequent resonance decays to stable hadrons.  We use the 
compilation of experimental data for particle number ratios in Au+Au 
at the BNL AGS (11$A$ GeV/$c$) and Pb+Pb at the CERN SPS (160$A$ 
GeV/$c$) which were presented by Stachel at QM'96 (see 
Ref.~\cite{stachel} and references therein).  Our fits within the VDW 
HG model is shown in Fig.~1 (see also \cite{yen97}).  From fitting 
the data we found the following model parameters: 
\begin{equation}\label{AGS}
\mbox{AGS}:~~~T~\cong~140~\mbox{MeV}~,~~\mu_b~\cong~590~
\mbox{MeV}~,
\end{equation}
\begin{equation}\label{SPS}
\mbox{SPS}:~~~T~\cong~185~\mbox{MeV}~,~~\mu_b~\cong~270~
\mbox{MeV}~.
\end{equation}
There are a number sets of possible values for hard-core radii 
which give the same hadron ratios in Fig.~1.  If we put $r_{\pi}=0$, 
then one finds $r=0.50$ fm and $r=0.46$ fm for AGS and SPS cases, 
respectively.  On the other hand, the AGS and SPS data in Fig.~1 can 
be fitted simultaneously with following set of 
\begin{equation}\label{core}
r_{\pi}~\cong~ 0.62~ \mbox{fm}~,~~~~ r~\cong ~0.8~ \mbox{fm}~.  
\end{equation}
Due to $r_{\pi}<r$ the ratios of {\it direct} pions to other hadrons 
increase by a factor of 2.0 for AGS (\ref{AGS}) and 2.6 for SPS 
(\ref{SPS}) in comparison to the case $r_{\pi}=r$ (in this later case 
all ratios are very close to their ideal gas values).  It results in 
the increase of {\it total} pion to other hadron ratios about 25$\%$ 
for AGS (\ref{AGS}) and 33$\%$ for SPS (\ref{SPS}).  

The calculated energy density $\varepsilon$ for the above $T$, 
$\mu_b$ values (\ref{AGS},\ref{SPS}) are 
\begin{equation}\label{AGS1}
\mbox{AGS}:~~~\varepsilon ^{id}~\cong~0.72~\mbox{GeV}/\mbox{fm}^3~,~~
\varepsilon ^{VDW}~=~0.25-0.08~\mbox{GeV}/\mbox{fm}^3~, 
\end{equation}
\begin{equation}\label{SPS1}
\mbox{SPS}:~~~\varepsilon ^{id}~\cong~1.58~\mbox{GeV}/\mbox{fm}^3~,~~
\varepsilon ^{VDW}~=~0.43-0.10~\mbox{GeV}/\mbox{fm}^3~, 
\end{equation}
for the ideal HG (i.e. $r_{\pi}=r=0$) and VDW HG, respectively.  The 
corresponding values for the baryonic number density $n_b$ are 
\begin{equation}\label{AGS2}
\mbox{AGS}:~~~n_b ^{id}~\cong~0.40~\mbox{fm}^{-3}~,~~n_b ^{VDW}~=~
0.13-0.04~\mbox{fm}^{-3}~,
\end{equation}
\begin{equation}\label{SPS2}
\mbox{SPS}:~~~n_b^{id}~\cong~0.40~\mbox{fm}^{-3}~,~~n_b ^{VDW}~=~
0.10-0.04~\mbox{fm}^{-3}~.  
\end{equation}
The higher values of $\varepsilon^{VDW}$ and $n_b ^{VDW}$ in the 
above equations are obtained for $r_{\pi}=0$, $r=0.50$~fm (AGS) and 
$r_{\pi}=0$, $r=0.46$ fm (SPS) while lower ones are found with 
parameters of Eq.~(\ref{core}) which fit simultaneously the AGS and 
SPS data.  Note also that Eq.~(3) gives an additional suppression by 
a factor of $\exp(-~v_i~p/T)$ for all particle number and energy 
densities of $i$-th particle in comparison to the inconsistent VDW 
gas treatment [7--10].  In the region of different $T$, $\mu_b$, 
$r_{\pi}$ and $r$ considered above it leads to a suppression factor 
between 2 and 6 for the total energy density and baryonic number 
density.  

To compare the energy densities of HG (\ref{AGS1},\ref{SPS1}) with 
those  of the QGP we use equation of state of the bag model, i.e., an 
ideal gas of $u,d,s$ quarks, antiquarks and gluons with 
nonperturbative ``vacuum pressure'', $ B~\cong~400$ MeV fm$^{-3}$ 
(see, e.g. \cite{bag}).  The chemical potentials of quarks and 
antiquarks are given by the general equation (\ref{mu}) with 
$b_i=1/3~ (-1/3)$ for all quarks (antiquarks), $s_i=0$ for $u,d$ 
quarks and antiquarks, $s_i=-1~ (+1)$ for $s$ quark (antiquark).  The 
consequence of zero strangeness requirement (\ref{strange}) is quite 
simple in the QGP.  It leads to equal values for the chemical 
potentials of $s$ and $\overline{s}$ and hence both should be equal 
to zero.  Thermodynamical functions within zero quark mass 
approximation can be easily calculated: 
\begin{equation}\label{pqgp}
p^{Q}~=~\frac{\pi^2}{90}~\frac{95}{2}~T^4~+~\frac{1}{9}~T^2~\mu_b^2~
+~\frac{1}{162\pi^2}~\mu_b^4~-~B~, 
\end{equation}
\begin{equation}\label{epsqgp}
\varepsilon^{Q}~
=~\frac{\pi^2}{30}~\frac{95}{2}~T^4~+~\frac{1}{3}~T^2~\mu_b^2~
+~\frac{1}{54\pi^2}~\mu_b^4~+~B~, 
\end{equation}
\begin{equation}\label{nbqgp}
n_b^Q~=~\frac{2}{9}~\mu_b~T^2~+~\frac{2}{81\pi^2}~\mu_b^3.  
\end{equation}

The lowest estimate of the energy density for the QGP can be 
obtained by putting 
\begin{equation}\label{pqgp0}
p^{Q}(T,\mu_b)~=~0~.  
\end{equation}
which defines a curve in the $\mu_b$-$T$ plane as shown in Fig.~2.  The 
values of $\varepsilon^Q$ (\ref{epsqgp}) ($\varepsilon ^Q=4B$) and 
$n_b^Q$ (\ref{nbqgp}) along this curve are shown in Fig.~3.  Note 
that the curve in the $\mu_b$-$T$ plane in Fig.~2 is not realistic for 
the HG-QGP phase transition.  This is just a lower estimate for the 
$\mu_b$-$T$ boundary of such a transition.  The values of $T$ and 
$\mu_b$ given in Eqs.~(\ref{AGS},\ref{SPS}) are also shown as points 
in Fig.~2: AGS (\ref{AGS}) square-point is very close to the QGP 
boundary and SPS~(\ref{SPS}) circle-point is in the ``QGP phase''.  
Our Fig.~2 is similar to the corresponding figures in 
Refs.~\cite{braun95b,stachel}.  To compare the energy densities of HG 
and QGP states one should remember that the energy density values are 
very different in the ideal HG and VDW HG.  The ideal HG model was 
used recently to describe particle number ratios in 
AGS~\cite{cleym97} and SPS~\cite{bec97} $A$+$A$ collisions.  Note 
that large values of $T\cong 193$ MeV and $\mu_b\cong 234$ MeV, close 
to our estimates (\ref{SPS}), were found in Ref.~\cite{bec97} for 
Pb+Pb (SPS).  Fig.~3 shows that ideal HG value of $\varepsilon^{id}$ 
for the SPS parameters (\ref{SPS}) is indeed rather close to the QGP 
phase.  However, the VDW HG values $\varepsilon^{VDW}$ for the same 
$T$ and $\mu_b$ are very much smaller than $\varepsilon^{id}$.  Ideal 
HG and VDW HG models lead to very different pictures of the chemical 
freezeout for $A$+$A$ collisions at SPS energies: 
$\varepsilon^{id} \cong \varepsilon^{Q}$, but 
$\varepsilon^{VDW}<<\varepsilon^{Q}$ and $n_b^{VDW}<<n_b^Q$.  We 
stress that very large values of $\varepsilon^{id}$ and	$n_b^{id}$ are 
in an inevitable logic contradiction with the assumed picture of 
non-interacting hadrons.  

In summary, we have analyzed the preliminary data for particle number 
ratios measured in the collisions of Au+Au at the BNL AGS 
(11$A$ GeV/$c$) and Pb+Pb at the CERN SPS (160$A$ GeV/$c$) with a 
thermodynamically consistent hadron gas model with excluded volume.  
Even though large values of temperature, $T=140$--185 MeV, and 
baryonic chemical potential, $\mu_b=590$--270 MeV, close to the 
boundary of the quark-gluon plasma phase are found from fitting the 
data, the energy densities obtained always lie far below a lowest 
limit estimate of the QGP energy density because of the VDW 
suppression effects.  Such a picture is consistent with a large 
energy density change during a HG-QGP phase transition.  

\acknowledgements
We thank M.~Ga\'zdzicki and C.~Spieles for discussions and critical 
comments.  MIG and HS are also thankful to J.~Eisenberg for the kind 
hospitality at the Tel Aviv University.  This work is supported in 
part by the BMFT, DFG, GSI and the NSC of Taiwan, ROC under grant 
numbers: 87-2112-M-001-005 and 87-2112-M-002-015.


\figure{ Fig.~1: 
Points are the preliminary experimental data for the particle number 
ratios (see Ref.~\cite{stachel} and references therein) for Au+Au AGS 
and Pb+Pb SPS collisions (in the lower and upper part of the figure 
respectively).  The short horizontal lines are the model fit with 
$T\cong 140$ MeV, $\mu_b \cong 590$ MeV (AGS), and $T\cong 185$ MeV, 
$\mu_b \cong 270$ MeV (SPS).  In both cases $r_{\pi}=0.62$~fm, 
$r=0.8$~fm.  }

\figure{ Fig.~2: 
The solid line is the curve $p^{Q}(T,\mu_b)=0$ (\ref{pqgp0}) in the 
$\mu_b$-$T$ plane. The square and circle are the chemical freeze-out 
points of (\ref{AGS}) and (\ref{SPS}), respectively.  }

\figure{ Fig.~3: 
The straight solid line gives the lowest energy density estimate, 
$\varepsilon ^Q=4B\cong 1.6$~GeV/fm$^3$, for the QGP.  It corresponds 
to the curve $p^{Q}(T,\mu_b)=0$ (\ref{pqgp0}) of Fig.~2 in the 
$n_b$-$\varepsilon$ plane.  The largest value in the $x$-axis, 
$n_b\cong 1.08$~fm$^{-3}$, corresponds to the point $T=0$ of the 
curve in Fig.~2.  The open square and circle show the ideal HG values 
of $\varepsilon^{id}$, $n_b^{id}$ for AGS (\ref{AGS1}),(\ref{AGS2}) 
and SPS (\ref{SPS1}),(\ref{SPS2}).  The solid lines show the sets of 
the VDW HG values $\varepsilon^{VDW}$, $n_b^{VDW}$ for AGS and SPS.  
The solid square and circle are the VDW HG values at AGS (\ref{AGS}) 
and SPS (\ref{SPS}) with hard-core radii (\ref{core}).  }

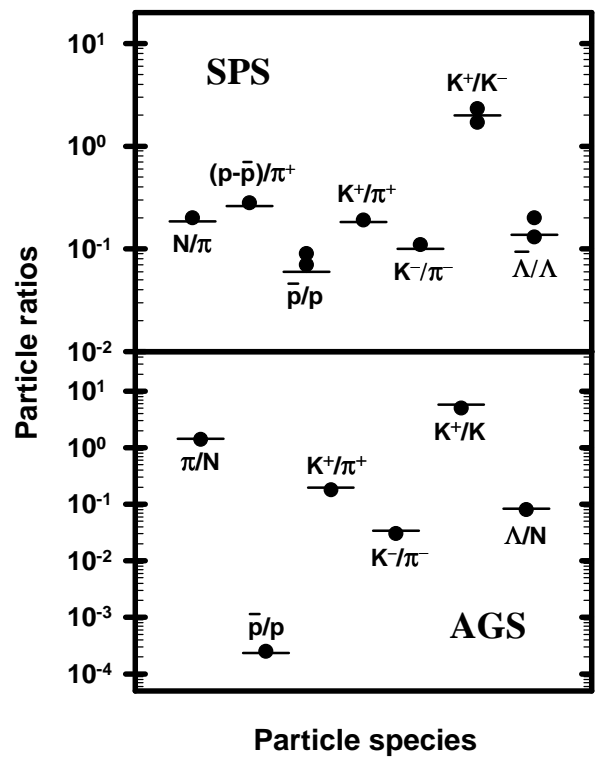
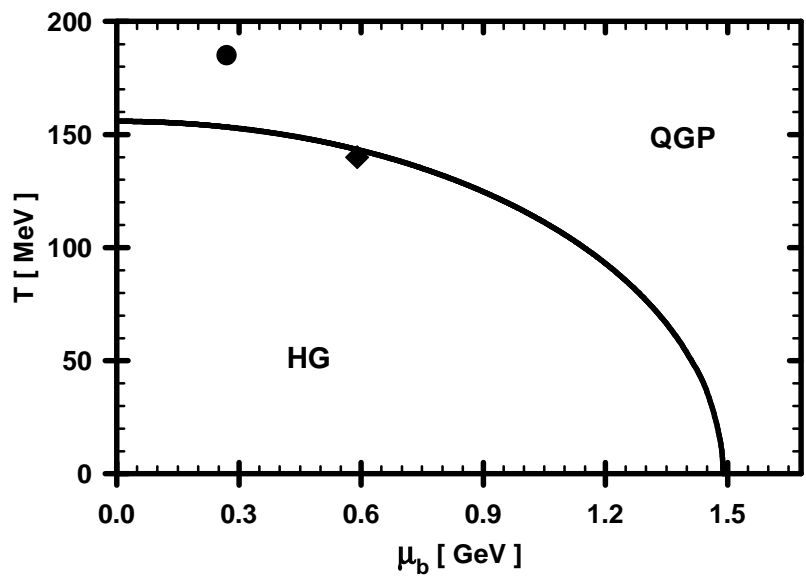
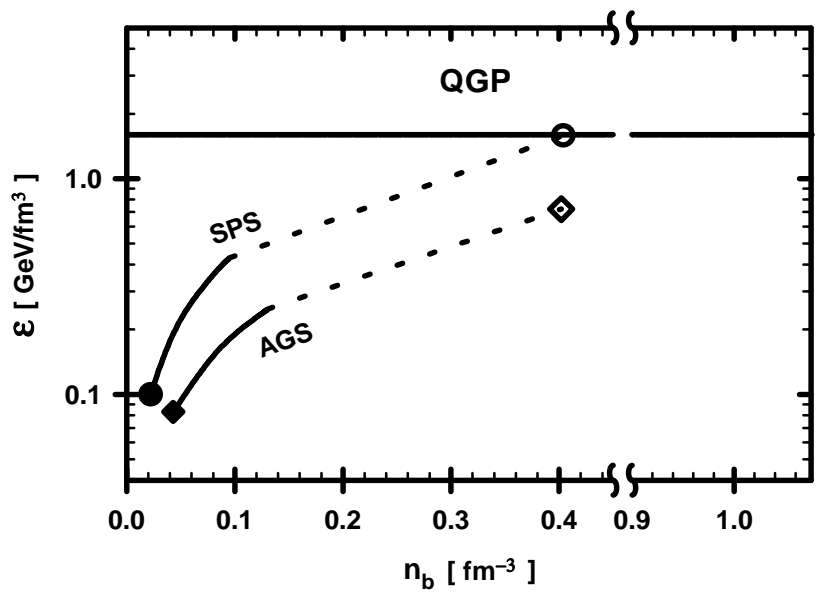
\end{document}